\documentclass[preprint,aps,showpacs,preprintnumbers,amsmath,amssymb]{revtex4}

\usepackage{epsfig}
\usepackage{graphicx}
\usepackage{subfigure}
\usepackage{dcolumn}
\usepackage{multirow}
\usepackage{bm}
\usepackage{verbatim}

\begin{document}


\title{Quantum tunneling-enhanced charging of nanoparticles in plasmas}

\author{Yu.~Tyshetskiy}
\email{y.tyshetskiy@physics.usyd.edu.au}
\author{S.~V.~Vladimirov}

\affiliation{%
School of Physics, University of Sydney 2006 NSW Australia
}%

\date{\today}

\begin{abstract}
The role of quantum tunneling effect in the electron accretion current onto a negatively charged grain immersed in isotropic plasma is analyzed, within the quasiclassic approximation, for different plasma electron distribution functions, plasma parameters, and grain sizes. It is shown that this contribution can be small (negligible) for relatively large (micron-sized) dust grains in plasmas with electron temperatures of the order of a few eV, but becomes important for nano-sized dust grains (tens to hundreds nm in diameter) in cold and ultracold plasmas (electron temperatures $\sim$ tens to hundreds of Kelvin), especially in plasmas with depleted high-energy \textquotedblleft tails\textquotedblright\ in the electron energy distribution.
\end{abstract}

\maketitle

Complex plasmas -- plasmas with dust particles (grains) in them~\cite{Vladimirov_Ostrikov_04,Fortov_UFN_04,Vlad_Book_05,Tsyt_book_08,Shukla_Eliasson_review} -- to a large extent owe their complexity to the fact that the charging process of a dust grain embedded in a plasma is sensitive to the plasma parameters and to proximity of other grains. Understanding the physics of grain charging is thus important for understanding complex plasmas.

The most commonly used model for finding the equilibrium charge of a grain immersed in plasma is the Orbital Motion Limited (OML) model~\cite{Allen_92,Goree_94}, in which the electron and ion currents from plasma onto the grain are found by analyzing particle orbits and determining whether they intersect the grain, using classical mechanics. The equilibrium grain charge is found from the condition that these classical currents cancel each other. However, in certain conditions quantum mechanical effects may become important, especially for electrons, and may lead to significant change of these currents, and hence of the equilibrium grain charge. Examples of additional currents induced by quantum effects include electron photoemission current (if the grain is illuminated by sufficiently energetic photons), or thermionic and/or field electron emission currents from the grain to the surrounding plasma~\cite{Raizer_book,Vlad_Book_05,Tsyt_book_08}. The latter currents are due to quantum tunneling of electrons from a negatively charged grain through the grain's potential barrier into the plasma, a process analogous to that responsible for alpha-decay of radioactive atoms~\cite{Gamow_28,Gurney_Condon_29}.

The processes of spontaneous and field-assisted tunneling of thermal electrons out of negatively charged grains into plasma, and the associated emission current densities, have been well studied in the literature~\cite{Guth_Mullin_42,Murphy_Good_56,Lee_59,Sodha_Kaw_68,Sodha_etal_09}. However, an inverse process of quantum tunneling of plasma electrons onto the negatively charged grain, which might, under favorable conditions, significantly increase the rate of electron accretion from plasma onto the grain, has not received proper attention, with an exception of Ref.~\cite{Middleton_04} where the cross-section for electron collisions with a spherical grain at low energies has been calculated quantum-mechanically.

The aim of this work is thus to calculate the additional current associated with quantum tunneling of plasma electrons, that are classically forbidden to overcome the repulsive potential barrier, onto the negatively charged grain. We compare this additional quantum tunneling current with the classical electron current from plasma onto the grain, and analyze how this additional current affects the self-consistent equilibrium grain charge, for different plasma parameters and grain sizes.

Consider a spherical grain immersed in an isotropic plasma with electrons and positive singly ionized ions. As the grain interacts with its environment, absorbing electrons and ions from the plasma and emitting electrons via processes such as photo and/or thermionic emission, it acquires an equilibrium net charge. Here, we assume this charge to be negative, which is normally the case due to higher mobility of plasma electrons compared to plasma ions, and relatively minor contribution of electron emission processes from plasma to the grain. [Note that a significant electron emission from the grain can make the total charge of the grain positive, in which case the electron and ion currents on the grain are found trivially from the classical OML theory. We will therefore not consider the case of positively charged grain here.]

To determine the electron accretion current onto the grain, consider the motion of electrons in the stationary central field of the spherical negatively charged grain in isotropic plasma. This motion is in general described by the stationary Schr{\"o}dinger's equation for the electron wave function $\psi$~\cite{L&L_v3}
\begin{equation}
\nabla^2\psi + \frac{2m_e}{\hbar^2}\left[E-U(r)\right]\psi = 0,
\end{equation}
where $E$ is the electron's total energy, $U(r)$ is the electron's potential energy in the central field. Seeking the solution $\psi$ in form of spherical harmonics $\psi(r,\theta,\varphi)=R(r)Y_{lm}(\theta,\varphi)$, we obtain an equation for the radial part $R(r)$ of the wave function, which can be easily reduced to the following equation for $\chi(r)=r R(r)$:
\begin{equation}
\frac{d^2\chi}{dr^2} + \left[\frac{2m_e}{\hbar^2}(E-U(r)) - \frac{l(l+1)}{r^2}\right]\chi = 0,  \label{eq:chi(r)}
\end{equation}
where $l$ is the angular momentum quantum number of the electron. This equation is equivalent to a one-dimensional Schr{\"o}dinger's equation with the effective potential energy
\begin{equation}
U_{\rm eff}(r) = U(r) + \frac{\hbar^2}{2m_e}\frac{l(l+1)}{r^2} = U(r) + \frac{J^2}{2m_e r^2},   \label{eq:U_eff_def}
\end{equation}
where $J^2 = \hbar^2 l(l+1)$ is the square of electron's angular momentum with respect to the center of the grain at $r=0$.

\begin{figure}
\includegraphics[width=3.2in]{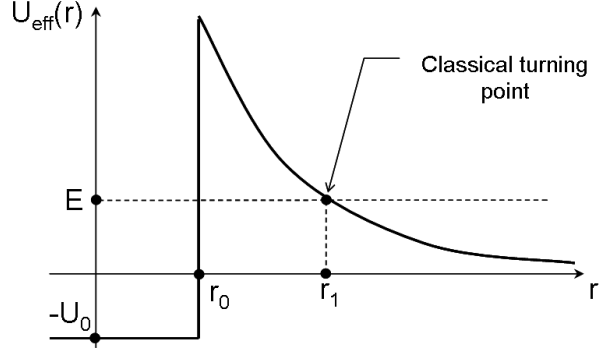}
\caption{\label{fig:radial_motion_sketch} A sketch of radial motion of an electron incoming from infinity with energy $E$ and angular momentum $J$ in the effective central potential (\ref{eq:U_eff_def}) of the grain of radius $r_0$.}
\end{figure}

In what follows, we consider the quasiclassical approximation of electron's motion in the effective potential $U_{\rm eff}(r)$, assuming that the electron's de Broglie wavelength is small compared to the characteristic scale of variation of $U_{\rm eff}(r)$ (the criterion of validity of the quasiclassical approximation will be discussed below). The radial motion of an electron of energy $E$ in the effective potential $U_{\rm eff}(r)$ is sketched in Fig.~\ref{fig:radial_motion_sketch}. An electron with total energy $E$ and absolute value of angular momentum $J=\hbar\sqrt{l(l+1)}$, coming towards the grain from infinity, will hit the grain of radius $r_0$ if $E\geq U_{\rm eff}(r_0)$. If $E < U_{\rm eff}(r_0)$, it encounters a potential barrier of the width $r_1 - r_0$, where $r_1$ is the classical turning point of the electron defined from $U_{\rm eff}(r_1) = E$. In classical mechanics, the electron cannot penetrate this barrier, and is reflected back to infinity, but in quantum mechanics the electron can tunnel through the barrier onto the grain with some non-zero probability $w_t$. In quasiclassical approximation, for the effective potential sketched in Fig.~\ref{fig:radial_motion_sketch}, this probability can be obtained in the form~\cite{L&L_v3}
\begin{equation}
w_t = \exp\left\{-\frac{2}{\hbar}\int_{r_0}^{r_1}{\sqrt{2 m_e\left[U_{\rm eff}(r) - E\right]} dr} \right\}.   \label{eq:w_t_gen}
\end{equation}

\begin{figure}
\includegraphics[width=3.2in]{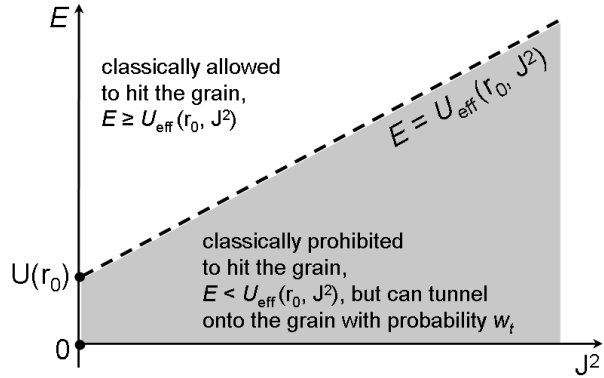}
\caption{\label{fig:EJ_plane} Mapping of electron orbits (allowed and not allowed to hit the grain) on the semiplane of the integrals of motion $E, J^2$.}
\end{figure}

Since the radial motion of an electron in a central field is fully defined by the two conserving quantities -- the electron's energy $E$ and angular momentum $\mathbf{J}$ with respect to the center of the field, one can map all possible trajectories of the electrons, incoming from infinity, onto a $\{E, J^2\}$ semiplane~\cite{B&R_59} shown in Fig.~\ref{fig:EJ_plane}, with the probability density of such trajectories defined by the distribution function of the incoming electrons expressed in terms of $E$ and $J^2$, $f_e^{(-)}(E,J^2)$ (here the superscript $(-)$ denotes the electrons with negative radial velocities at infinity). On this semiplane, the line $E = U_{\rm eff}(r_0, J^2)$ corresponds to the minimum energy $E$ that an electron with angular momentum $J$ needs to have, in order to be allowed by classical mechanics to hit the grain. In other words, this line separates the regions of parameters $E,J^2$ for which the electrons are allowed (above the line) or prohibited (below the line) by the classical mechanics to hit the grain. The corresponding \textquotedblleft classical\textquotedblright\ electron current onto the grain, i.e., the current due to electrons with $E\geq U_{\rm eff}(r_0, J^2)$ that are classically allowed to hit the grain, is~\cite{B&R_59}
\begin{equation}
I_{\rm clas}^{(e)} = 4\pi r_0^2 \frac{\pi}{m_e^3 r_0^2} \iint_{E\geq U_{\rm eff}(r_0,J^2)}{dE dJ^2\ f_e^{(-)}(E,J^2)\cdot 1},  \label{eq:I_e_clas}
\end{equation}
where $1$ in the integral stands for the probability for an electron to hit the grain, which for electrons with energies $E\geq U_{\rm eff}(r_0, J^2)$ is equal to unity. The current on the grain due to electrons with $E < U_{\rm eff}(r_0, J^2)$, which we call the \textquotedblleft tunneling\textquotedblright\ current, is then
\begin{equation}
I_{\rm tun}^{(e)} = 4\pi r_0^2 \frac{\pi}{m_e^3 r_0^2} \iint_{E < U_{\rm eff}(r_0,J^2)}{dE dJ^2\ f_e^{(-)}(E,J^2) \cdot w_t(E,J^2)}, \label{eq:I_tun_int}
\end{equation}
with the tunneling probability $w_t(E,J^2)$ defined by (\ref{eq:w_t_gen}) in the quasiclassical approximation for electron radial motion. Note that in classical mechanics, the tunneling probability for electrons with $E < U_{\rm eff}$ is zero [one can see this by formally taking the limit $\hbar\rightarrow 0$ in (\ref{eq:w_t_gen})], and thus $I_{\rm tun}^{(e)}=0$, as expected. With $J^2 = \hbar^2 l(l+1)$ we have $dJ^2=\hbar^2(2l+1)$, and the integration over $J^2$ in (\ref{eq:I_tun_int}) turns into a sum over $l$:
\begin{equation}
I_{\rm tun}^{(e)} = 4\pi r_0^2 \frac{\pi}{m_e^3 r_0^2} \sum_{l=0}^\infty{\hbar^2 (2l+1)\int_{0}^{U_{\rm eff}(r_0,l)}{dE\ f_e^{(-)}(E,l) \cdot w_t(E,l)}}. \label{eq:I_tun_sum}
\end{equation}
where
\begin{equation}
U_{\rm eff}(r_0,l) = U(r_0) + \frac{\hbar^2}{2 m_e}\frac{l(l+1)}{r_0^2}.
\end{equation}

\begin{figure}
\includegraphics[width=3.2in]{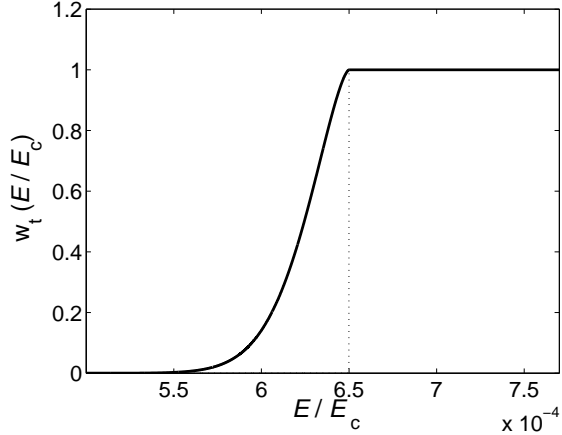}
\caption{\label{fig:w_t(E)} Dependence of tunneling probability $w_t$ on the normalized electron energy $\tilde{E}=E/E_c$ (where $E_c=\alpha^2 m_e/\hbar^2$ is the energy in Coulomb units), for a fixed value of $l>0$.}
\end{figure}

The tunneling probability $w_t(E,J^2)$ can be evaluated analytically for a Coulomb (i.e., unscreened by plasma) potential of the grain, $U(r) = \alpha/r$, with $\alpha=Z_d e^2$, where $Z_d$ is the charge of the grain in electron charges. [Note that by assuming the unscreened grain potential, we somewhat overestimate the width of the barrier through which an electron with given $J$ and $E<U_{\rm eff}$ has to tunnel to reach the grain, compared to the real screened grain potential (e.g., the Debye-H{\"u}ckel potential). As a result, we expect to underestimate the tunneling probability, and thus the tunneling current obtained below for the unscreened grain is expected to be somewhat less than the tunneling current on a grain screened by plasma.] Introducing Coulomb units of mass, length, time, and energy as $m_e$, $\hbar^2/(\alpha m_e)$, $\hbar^3/(\alpha^2 m_e)$ and $\alpha^2 m_e/\hbar^2$, respectively, we define the corresponding dimensionless quantities as
\begin{equation}
\tilde{m} = \frac{m}{m_e},\ \ \ \tilde{r} = \frac{m_e\alpha}{\hbar^2}r,\ \ \ \tilde{E} = \frac{\hbar^2}{m_e\alpha^2}E,\ \ \ \tilde{n}_e = \left(\frac{\hbar^2}{m_e\alpha}\right)^2 n_e.
\end{equation}
Then, from (\ref{eq:I_tun_sum}) we have for the tunneling electron current from plasma on an unscreened negatively charged grain of radius $r_0$:
\begin{equation}
I_{\rm tun}^{(e)} = \frac{4\pi^2\alpha^2}{\tilde{m}^2\hbar^2} \sum_{l=0}^\infty{\hbar^2 (2l+1)\int_{0}^{\tilde{U}_{\rm eff}(\tilde{r}_0,l)}{d\tilde{E}\ w_t(\tilde{E},l)}} f_e^{(-)}(\tilde{E},l), \label{eq:I_tun_sum_dimless}
\end{equation}
where
\begin{eqnarray}
\tilde{U}_{\rm eff}(\tilde{r}_0,l) &=& \frac{1}{\tilde{r}_0} + \frac{l(l+1)}{2\tilde{r}_0^2}, \label{eq:U_eff} \\
w_t(\tilde{E},l) &=& \exp\left\{-\sqrt{8}\int_{\tilde{r}_0}^{\tilde{r}_1}{\left(\frac{1}{\tilde{r}} + \frac{l(l+1)}{2\tilde{r}^2} - \tilde{E}\right)^{1/2} d\tilde{r}} \right\},  \label{eq:w_t_dimless}
\end{eqnarray}
with the dimensionless classical turning point $\tilde{r}_1$ of an electron with $E<U_{\rm eff}$ given by
\begin{equation}
\tilde{r}_1(\tilde{E},l) = \frac{1}{2\tilde{E}}\left(1+\sqrt{1+2\tilde{E}\ l(l+1)}\right).
\end{equation}
To evaluate $w_t(\tilde{E},l)$ from (\ref{eq:w_t_dimless}), we use the formulas~\cite{G&R_p97}:
\begin{eqnarray}
\int{\frac{\sqrt{R}}{x^2}dx} &=& -\frac{\sqrt{R}}{x} + \frac{b}{2}\int{\frac{dx}{x\sqrt{R}}} + c\int{\frac{dx}{\sqrt{R}}},\ \nonumber \\
&&\text{where }\ \ R = a + bx + cx^2,\ \ \text{for }a\neq 0, \Delta = 4ac-b^2<0,\\
\int{\frac{\sqrt{bx + cx^2}}{x^2}dx} &=& -\frac{2\sqrt{bx+cx^2}}{x} + c\int{\frac{dx}{\sqrt{bx+cx^2}}},\ \ \ \text{for }a=0,
\end{eqnarray}
with $a=-\tilde{E}\leq 0$, $b=1$, $c=l(l+1)/2$, and $\Delta=-(1+2\tilde{E}\ l(l+1))<0$. Performing the integrations, we obtain for $w_t(\tilde{E},l)$ in case of unscreened grain potential:
\begin{eqnarray}
&&w_t(\tilde{E},l) = \exp[-\sqrt{8}(L_2-L_1)], \label{eq:w_t_Coulomb} \\
&&L_1 = \tilde{r}_0\sqrt{\frac{l(l+1)}{2\tilde{r}_0^2} + \frac{1}{\tilde{r}_0} - \tilde{E}} - \frac{1}{2\tilde{E}}\arcsin\left(\frac{1-2\tilde{E}\tilde{r}_0}{\sqrt{1+2\tilde{E}\ l(l+1)}}\right) \nonumber \\
&& - \sqrt{\frac{l(l+1)}{2}}\ln\left(\frac{1+l(l+1)/\tilde{r}_0 + \sqrt{2l(l+1)}\sqrt{\tilde{E} - l(l+1)/2\tilde{r}_0^2 - 1/\tilde{r}_0}}{\sqrt{1 + 2\tilde{E}\ l(l+1)}}\right), \\
&&L_2 = \frac{\pi}{4\sqrt{\tilde{E}}} + \frac{i\pi}{2}\sqrt{\frac{l(l+1)}{2}}.
\end{eqnarray}
[Note that $w_t$ is a real quantity, as the imaginary parts of $L_1$ and $L_2$ cancel each other.] A representative plot of $w_t(\tilde{E},l)$ for a fixed $l>0$ is shown in Fig.~\ref{fig:w_t(E)}. The probability of electron tunneling from the classical turning point $r_1$ onto the grain increases with $\tilde{E}$ until it reaches the value of $1$ at $\tilde{E}=\tilde{U}_{\rm eff}(\tilde{r}_0,l)$ when the electron is classically allowed to hit the grain.

The classical (\ref{eq:I_e_clas}) and the tunneling (\ref{eq:I_tun_sum_dimless}) electron currents make up the total electron current from plasma onto the grain, which charges the grain negatively. On the other hand, accretion of positive ions produces the ion current onto the grain which charges the grain positively. Since the ions are heavy and are attracted by the negatively charged grain, the ion current on the grain is a classical one, and is defined by the OML theory, which for the Maxwell-Boltzmann distribution of ions yields~\cite{Fortov_UFN_04}
\begin{equation}
I^{(i)} = \sqrt{8\pi}r_0^2 n_i v_{Ti} (1-e\phi_s/T_i), \label{eq:I_i_clas}
\end{equation}
where $r_0$ is the grain radius, $n_i$ is the ion density far from the grain, $v_{Ti}=(T_i/m_i)^{1/2}$ is the ion thermal velocity, $T_i$ is the ion temperature in units of energy, and $\phi_s=\alpha/r_0$ is the surface potential of the grain. The balance of total electron and ion currents onto the grain defines the equilibrium surface potential $\phi_s=\alpha/r_0$ of the grain. The corresponding equation of balance of currents
\begin{equation}
I_{\rm clas}^{(e)} + I_{\rm tun}^{(e)} = I^{(i)}   \label{eq:I_e=I_i}
\end{equation}
is a nonlinear equation for the surface potential of the grain (and hence for the grain charge $Z_d$ since $\phi_s=\alpha/r_0$ with $\alpha = e^2 Z_d$), that can be solved iteratively, for a given distribution function $f_e^{(-)}$ of incoming plasma electrons far from the grain.

Below we calculate the ratio $I_{\rm tun}^{(e)}/I_{\rm clas}^{(e)}$ of electron tunneling and classical currents onto the grain, and the ratio $Z_d^{\rm clas+tun}/Z_d^{\rm clas}$ of grain equilibrium charges defined from (\ref{eq:I_e=I_i}) with and without the tunneling electron current, for several types of electron energy distributions and a range of plasma parameters and grain sizes.

\subsubsection{Maxwell-Boltzmann distribution of electrons}
For Maxwell-Boltzmann distribution of incoming electrons, $f_e^{(-)}(E,J^2) = f_M(E) = n_{0e}(m_e/2\pi T_e)^{3/2}\exp(-E/T_e)$, where $n_{0e}$ is the plasma electron density far from the grain, $T_e$ is the electron temperature in units of energy, we have for the classical and tunneling electron currents onto the grain:
\begin{eqnarray}
I_{\rm clas, M}^{(e)} &=& \sqrt{8\pi}r_0^2 n_{0e} v_{Te} \exp\left(-\frac{\alpha}{r_0 T_e}\right),\ \ \ \alpha=e^2 Z_d>0, \\
I_{\rm tun, M}^{(e)} &=& \frac{4\pi^2 n_{0e}}{\sqrt{m_e}}\frac{\alpha^2}{(2\pi T_e)^{3/2}}\sum_{l=0}^\infty{(2l+1)\int_0^{\tilde{U}_{\rm eff}(\tilde{r}_0,l)}{d\tilde{E}\ w_t(\tilde{E},l)\exp(-\tilde{E}/T_e)}},
\end{eqnarray}
with $\tilde{U}_{\rm eff}$ and $w_t$ defined by Eqs~(\ref{eq:U_eff}) and (\ref{eq:w_t_Coulomb}), respectively.

\subsubsection{Druyvesteyn distribution of electrons}
Although in theoretical calculations the Maxwellian distribution is usually assumed, the actual plasma electron distribution function is often significantly non-Maxwellian, such as in gas disharges~\cite{Tsendin_rev_09}. For example, experimental measurements in inductively coupled plasma (ICP)~\cite{Godyak_Piejak_PRL_90} and capacitively coupled plasma (CCP)~\cite{Barnes_etal_APL_93} discharges revealed that, while at low pressures the electron distribution function is indeed close to Maxwellian, at higher pressures ($p\gtrsim 10$~mTorr in ICP, and $p\gtrsim 0.5$~Torr in CPP) it turns into a Druyvesteyn-like distribution, with depleted high-energy tail. Obviously, this should lead to a change in the electron classical and tunneling currents on the grain, and consequently in the equilibrium grain charge, compared to the case of Maxwellian electron distribution.

For Druyvesteyn distribution of incoming electrons, $f_e^{(-)}(E,J^2) = f_D(E) = n_{0e}A_D(m_e/\pi T_e)^{3/2}\exp[-B_D(E/T_e)^2]$, where the constants $A_D\approx 0.177$ and $B_D\approx 0.243$ are defined from the conditions $\left<n_e\right>=n_{0e}$ and $\left<E\right>=(3/2) T_e$ (here $\left< ...\right>$ denotes the average over all electrons), we have for the classical and tunneling electron currents onto the grain:
\begin{eqnarray}
I_{\rm clas, D}^{(e)} &\approx& 0.25\sqrt{8\pi} r_0^2 n_{0e} v_{Te} \left\{\frac{1}{B_D}\exp\left[-B_D\left(\frac{\alpha}{r_0 T_e}\right)^2\right]\right. \nonumber \\
&-&\left.\sqrt{\frac{\pi}{B_D}}\frac{\alpha}{r_0T_e}\left[1-{\rm Erf}\left(\sqrt{B_D}\frac{\alpha}{r_0 T_e}\right)\right] \right\}, \\
I_{\rm tun, D}^{(e)} &\approx& 0.50\frac{4\pi^2 n_{0e}}{\sqrt{m_e}}\frac{\alpha^2}{(2\pi T_e)^{3/2}}\sum_{l=0}^\infty{(2l+1)\int_0^{\tilde{U}_{\rm eff}(\tilde{r}_0,l)}{d\tilde{E}\ w_t(\tilde{E},l)\exp\left[-B_D\left(\frac{\tilde{E}}{T_e}\right)^2\right]}},
\end{eqnarray}
where $\rm Erf$ is the error function, and $\tilde{U}_{\rm eff}$ and $w_t$ are defined by Eqs~(\ref{eq:U_eff}) and (\ref{eq:w_t_Coulomb}), respectively.

\subsubsection{Step distribution of electrons}
In complex plasmas, dust grains absorb electrons with energies above $U_{\rm eff}(r_0,J^2)$, which depletes the high-energy \textquotedblleft tail\textquotedblright\ of the electron distribution function \cite{Tsyt_book_08}. For high enough number density of the grain component, this effect becomes significant, and the resulting electron distribution function can be roughly approximated by the step function
\begin{equation}
f_e^{(-)}(E,J^2) = f_S(E) = \left\{
\begin{matrix}
(3\sqrt{2}/16\pi)n_{0e}(m_e/E_{\rm max})^{3/2}\ \ \ \text{for }E\leq E_{\rm max}, \\
0\ \ \ \ \ \ \ \ \ \ \ \ \ \ \ \ \ \ \ \ \ \ \ \ \ \ \ \ \ \ \ \ \ \ \ \ \ \text{for }E>E_{\rm max}.
\end{matrix}
\right.
\end{equation}
Here, $E_{\rm max} = e|\phi_s| + \delta E$, where $e|\phi_s|$ is the minimum energy at which an electron with $l=0$ (zero angular momentum with respect to the grain) hits the grain, and $\delta E$ accounts for the fact that most electrons have $l>0$. For simplicity, $\delta E$ can be approximated by a constant, defined, e.g., from experimental data for a particular complex plasma system. Here for definitiveness we will assume a reasonable value $\delta E = 0.5 T_e$. For the step distribution of incoming electrons, we have for the classical and tunneling electron currents onto the grain:
\begin{eqnarray}
I_{\rm clas, S}^{(e)} &=& \frac{3\pi}{2\sqrt{2 m_e}}r_0^2 n_{0e}\frac{\delta E^2}{\left(\alpha/r_0 + \delta E\right)^{3/2}}, \\
I_{\rm tun, S}^{(e)} &=& \frac{3\pi}{4}\sqrt{\frac{2}{m_e}}\frac{n_{0e}\alpha^2}{E_{\rm max}^{3/2}}\sum_{l=0}^\infty{(2l+1)\int_0^{\tilde{U}_{\rm eff}(\tilde{r}_0,l)}{d\tilde{E}\ w_t(\tilde{E},l) \sigma_{H}(\tilde{E}_{\rm max}-\tilde{E})}},
\end{eqnarray}
where $\sigma_H$ is the Heaviside step function, and $\tilde{U}_{\rm eff}$ and $w_t$ are defined by Eqs~(\ref{eq:U_eff}) and (\ref{eq:w_t_Coulomb}), respectively.

\begin{figure}
\includegraphics[width=3.2in]{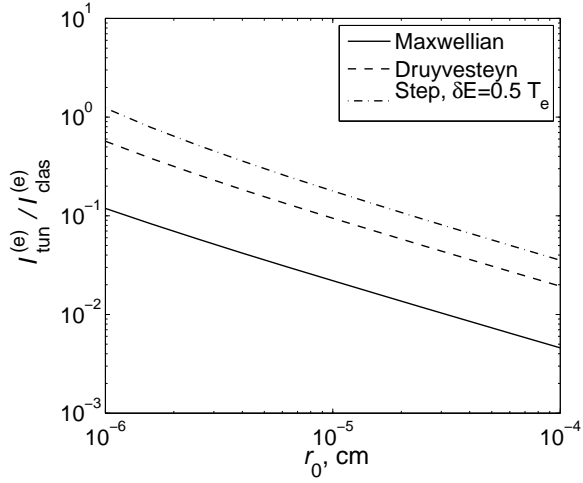}
\caption{\label{fig:eta_I_MDS} Comparison of $\eta_I=I_{\rm tun}^{(e)}/I_{\rm clas}^{(e)}$ as a function of the grain radius $r_0$, for different electron distribution functions: Maxwellian (solid line), Druyvesteyn (dashed line), and the step distribution with $\delta E = 0.5 T_e$ (dash-dotted line). The currents are calculated for grains with corresponding equilibrium charges, immersed in argon plasma with $T_e=0.1$~eV and $T_e/T_i=10^2$.}
\end{figure}

\begin{table}
\begin{tabular}{|c|c|c|c|c|c|c|c|c|c|}
\hline
 & \multicolumn{3}{|c|}{Maxwellian} & \multicolumn{3}{|c|}{Druyvesteyn} & \multicolumn{3}{|c|}{Step distribution}\\
\hline\hline
$I_{\rm tun}^{(e)}/I_{\rm clas}^{(e)}$&
\multicolumn{3}{c|}{Grain radius $r_0$, cm} & \multicolumn{3}{c|}{Grain radius $r_0$, cm} & \multicolumn{3}{c|}{Grain radius $r_0$, cm} \\
  & $10^{-6}$ & $10^{-5}$ & $10^{-4}$ & $10^{-6}$ & $10^{-5}$ & $10^{-4}$ & $10^{-6}$ & $10^{-5}$ & $10^{-4}$ \\ \hline
$T_e=0.01\ \text{eV}$ & $\ \sim 0.33\ $ & $\ \sim 0.05\ $ & $\ \sim 0.01\ $ & $\ \sim 2.0\ $ & $\ \sim 0.22\ $ & $\ \sim 0.04\ $ & $\ \sim 5.0\ $ & $\ \sim 0.43\ $ & $\ \sim 0.08\ $ \\ \hline
$T_e=0.1\ \text{eV}$ & $\sim 0.12$ & $\sim 0.02$ & $\sim 0.005$ & $\sim 0.57$ & $\sim 0.10$ & $\sim 0.02$ & $\sim 1.24$ & $\sim 0.18$ & $\sim 0.04$ \\ \hline
$T_e=1\ \text{eV}$ & $\sim 0.05$ & $\sim 0.01$ & $\sim 0.002$ & $\sim 0.22$ & $\sim 0.04$ & $\sim 0.01$ & $\sim 0.43$ & $\sim 0.08$ & $\sim 0.02$ \\
\hline
\end{tabular}
\caption{Ratio $I_{\rm tun}^{(e)}/I_{\rm clas}^{(e)}$ of tunneling and classical electron accretion currents onto the grain, for different electron distribution functions, electron temperatures, and grain sizes. The currents are calculated for grains with corresponding equilibrium charges, defined by Eq.~(\ref{eq:I_e=I_i}), immersed in argon plasma with $T_e/T_i=10^2$. \label{tab:I_ratio}}
\end{table}
\begin{table}
\begin{tabular}{|c|c|c|c|c|c|c|c|c|c|}
\hline
 & \multicolumn{3}{|c|}{Maxwellian} & \multicolumn{3}{|c|}{Druyvesteyn} & \multicolumn{3}{|c|}{Step distribution}\\
\hline\hline
$Z_d^{\rm clas+tun}/Z_d^{\rm clas}$&
\multicolumn{3}{c|}{Grain radius $r_0$, cm} & \multicolumn{3}{c|}{Grain radius $r_0$, cm} & \multicolumn{3}{c|}{Grain radius $r_0$, cm} \\
  & $10^{-6}$ & $10^{-5}$ & $10^{-4}$ & $10^{-6}$ & $10^{-5}$ & $10^{-4}$ & $10^{-6}$ & $10^{-5}$ & $10^{-4}$ \\ \hline
$T_e=0.01\ \text{eV}$ & $\ \sim 0.1\ $ & $\ \sim 0.02\ $ & $\ \sim 0.003\ $ & $\ \sim 0.25\ $ & $\ \sim 0.05\ $ & $\ \sim 0.01\ $ & $\ \sim 1.25\ $ & $\ \sim 0.18\ $ & $\ \sim 0.04\ $ \\ \hline
$T_e=0.1\ \text{eV}$ & $\sim 0.03$ & $\sim 0.007$ & $\sim 0.002$ & $\sim 0.1$ & $\sim 0.02$ & $\sim 0.005$ & $\sim 0.45$ & $\sim 0.08$ & $\sim 0.02$ \\ \hline
$T_e=1\ \text{eV}$ & $\sim 0.02$ & $\sim 0.003$ & $< 0.001$ & $\sim 0.05$ & $\sim 0.01$ & $\sim 0.002$ & $\sim 0.18$ & $\sim 0.04$ & $\sim 0.008$ \\
\hline
\end{tabular}
\caption{Ratio $Z_d^{\rm clas+tun}/Z_d^{\rm clas}$ of equilibrium grain charges defined by the current balance~(\ref{eq:I_e=I_i}) with and without the electron tunneling current, for different electron distribution functions, electron temperatures, and grain sizes. The charges are calculated for grains in argon plasma with $T_e/T_i=10^2$. \label{tab:Zd_ratio}}
\end{table}

The importance of the electron tunneling current onto the grain can be characterized by the ratio of the tunneling and classical electron currents $\eta_I=I_{\rm tun}^{(e)}/I_{\rm clas}^{(e)}$. The tunneling electron accretion current increases the total electron current onto the grain, thus offsetting the total electron-ion current balance, and changing the equilibrium grain charge. This change is characterized by the ratio $\eta_Z=Z_d^{\rm clas+tun}/Z_d^{\rm clas}$ of the grain charges defined from (\ref{eq:I_e=I_i}) with and without accounting for the electron tunneling current. The ratios $\eta_I$ and $\eta_Z$ depend on several factors: the electron distribution function, electron and ion temperatures, the grain size, and the type of gas used in the plasma discharge. These dependencies are illustrated in Fig.~\ref{fig:eta_I_MDS} and in Tables~\ref{tab:I_ratio}~and~\ref{tab:Zd_ratio}. Generally, since $\eta_I$ is proportional to the ratio of populations of electrons with $E\geq U_{\rm eff}(r_0,J^2)$ and $E< U_{\rm eff}(r_0,J^2)$, both $\eta_I$ and $\eta_Z$ increase for electron distributions with depleted tails, such as the Druyvesteyn or the step distribution, as seen in Fig.~\ref{fig:eta_I_MDS}. Due to the same reason, for a given type of electron distribution function, $\eta_I$ and $\eta_Z$ increase at lower electron temperatures $T_e$ (for a fixed ion temperature $T_i$), and increase for smaller grain sizes $r_0$, as seen from Fig.~\ref{fig:eta_I_MDS} and Tables~\ref{tab:I_ratio}~and~\ref{tab:Zd_ratio}. [Remarkably, since the ratio of populations of electrons with $E\geq U_{\rm eff}(r_0,J^2)$ and $E< U_{\rm eff}(r_0,J^2)$ increases with the slope of the $E=U_{\rm eff}$ line in Fig.~\ref{fig:EJ_plane} (i.e., decreases with the grain size $r_0$), the dependence of $\eta_I$ on the grain size $r_0$ can be well approximated by a universal power law, $\eta_I \propto r_0^{-0.7}$, where the coefficient of proportionality only depends on the plasma parameters (i.e., electron distribution function, electron and ion temperatures, gas atomic mass), and is (almost) independent of $r_0$.]

We should note that actually the electron distribution function almost never has a clear-cut form such as Maxwellian or Druyvesteyn or step function, but instead can be a superposition of several distributions of different type with different effective temperatures. Hence, the results of calculations of $\eta_I$ and $\eta_Z$ shown in Fig.~\ref{fig:eta_I_MDS} and in Tables~\ref{tab:I_ratio}-\ref{tab:Zd_ratio} should be perceived as illustrating the trends. But if the electron distribution function can be measured, the corresponding electron current onto the grain can be calculated using Eqs~(\ref{eq:I_e_clas}) and (\ref{eq:I_tun_sum}) with this distribution.

The criterion of validity of the quasiclassical approximation, in which the expression (\ref{eq:w_t_gen}) for the tunneling probability is obtained, reduces to the requirement that the electron's de Broglie wavelength $\lambdabar\sim\hbar/\sqrt{2 m_e |E|}$ is small compared to the size $\alpha/|E|$ of the region near the grain where the electron energy $E$ is of the order of $U_{\rm eff}$~\cite{L&L_v3}. Because $U_{\rm eff}$ is minimum for the electrons with $l=0$, this criterion is the strongest for such electrons, i.e., if it is satisfied for electrons with $l=0$, it is automatically satisfied for electrons with $l>0$. For $l=0$, this criterion reduces to $\hbar v/\alpha\ll 1$, where $v\sim\sqrt{E/m_e}$ is the classical electron velocity. With $v\sim v_{Te}=\sqrt{T_e/m_e}$ and $\alpha\sim r_0 z T_e$, where $z=e|\phi_s|/T_e$ is the dimensionless grain charge, $z\sim 1$, we have the following requirement for the grain size $r_0$ for which the quasiclassical approximation of electron motion is applicable:
\begin{equation}
r_0 \gg \frac{\hbar}{\sqrt{m_e T_e}}.   \label{eq:criterion}
\end{equation}
For $T_e\sim 1$ eV this gives $r_0\gg 10^{-8}$~cm, and for $T_e\sim 0.01$ eV this gives $r_0\gg 3\cdot10^{-7}$~cm, which is well satisfied for the grain sizes used in Fig.~\ref{fig:eta_I_MDS} and Tables~\ref{tab:I_ratio}~and~\ref{tab:Zd_ratio}. We note that for very small grains, or for very low electron temperatures, when the quasiclassical approximation becomes invalid, the Schr{\"o}dinger's equation~(\ref{eq:chi(r)}) for the radial wave function of an electron has to be solved exactly, in order to define the electron current from plasma to the grain. This is however hardly necessary, as the criterion~(\ref{eq:criterion}) is well satisfied for a wide range of plasma electron temperatures and grain sizes.

As seen from Fig.~\ref{fig:eta_I_MDS} and Tables~\ref{tab:I_ratio}~and~\ref{tab:Zd_ratio}, the effect of electron tunneling from plasma onto the grain is most pronounced and significant for small grains ($r_0\sim$ tens to hundreds of nm) in plasmas with low electron temperatures ($T_e\sim$ tens to hundreds of K), especially for electron distributions with depleted high-energy \textquotedblleft tails\textquotedblright. Note that we underestimated the tunneling current by assuming the grain field to be unscreened. In reality the grain is shielded by plasma and its potential decays faster than the Coulomb potential, narrowing the width of the potential barrier around the grain, and thus further increasing the electron tunneling probability and the electron accretion current onto the grain, especially in low temperature plasmas with small Debye lengths. We therefore expect the considered effect of quantum tunneling of plasma electrons onto the grain to be important in plasmas with nano and submicron size dust grains~\cite{Vladimirov_Ostrikov_04,Vlad_Book_05}, in ultracold plasmas (where the electron temperature can be as low as $30$~K~\cite{Roberts_etal_PRL_04}), and in dark molecular clouds in astrophysics~~\cite{Middleton_04,Pandey_Vlad_07}.

The authors thank A.~Samarian for useful discussions. This work was supported by the Australian Research Council.


\end{document}